\documentclass[11pt,onecolumn]{IEEEtran}
\usepackage{amsthm}
\usepackage{times,amssymb,amsmath,amsfonts,float,nicefrac,color,bbm,mathrsfs,caption,float}
\usepackage{algorithm,enumerate,multirow,caption,tikz,graphicx}
\usepackage[mathscr]{eucal}
\usepackage{sidecap}
\usepackage{algpseudocode}
\usepackage{verbatim}
\usepackage{epstopdf}
\usepackage{textcomp}
\usetikzlibrary{shapes,arrows}
\usepackage{stmaryrd}
\usepackage{mathabx}
\usepackage[noadjust]{cite}
\usepackage{booktabs}
\usepackage[top=1in,bottom=1in,left=1in,right=1in]{geometry}
\allowdisplaybreaks

\newcommand{\RN}[1]{%
  \textup{\expandafter{\romannumeral#1}}%
}

\newcommand\remove[1]{}

\allowdisplaybreaks

\newtheorem{theorem}{Theorem}
\newtheorem{definition}{Definition}
\newtheorem{proposition}{Proposition}

\newtheorem{claim}{Claim}
\newtheorem{lemma}{Lemma}

\newtheorem{example}{Example}
\newtheorem{remark}{Remark}

\newcommand{\ff}{{\mathbb F}}

\newcommand{\cC}{\mathcal{C}}

\newcommand{\cJ}{\mathcal{J}}

\newcommand{\cP}{\mathcal{P}}

\newcommand{\cR}{\mathcal{R}}
\newcommand{\cS}{\mathcal{S}}

\DeclareMathOperator{\trace}{tr}

\DeclareMathOperator{\lcm}{lcm}

\DeclareMathOperator{\Diag}{Diag}
\DeclareMathOperator{\spun}{Span}

\begin{document}
\title{Optimal repair of Reed-Solomon codes: Achieving the cut-set bound}

\author{\IEEEauthorblockN{Itzhak Tamo} \hspace*{1in}
\and \IEEEauthorblockN{Min Ye} \hspace*{1in}
\and \IEEEauthorblockN{Alexander Barg}}

\maketitle
{\renewcommand{\thefootnote}{}\footnotetext{

\vspace{-.2in}
 
\noindent\rule{1.5in}{.4pt}

{I. Tamo is with Department of EE-Systems, Tel Aviv University, Tel Aviv,
Israel. Email: zactamo@gmail.com.
His research is supported by ISF grant no.~1030/15 and the NSF-BSF grant no.~2015814.

M. Ye and A. Barg are with Dept. of ECE and ISR, University of Maryland, College Park, MD 20742. Emails: yeemmi@gmail.com and abarg@umd.edu. Their research is supported by NSF grants CCF1422955 and CCF1618603.}}
}
\renewcommand{\thefootnote}{\arabic{footnote}}
\setcounter{footnote}{0}

\begin{abstract}
In distributed storage systems, data is written on a large number of individual storage nodes. The data is stored in encoded form to protect it from node failures, and the coding method of choice relies on Maximum Distance Separable (MDS) codes. Individual coordinates of the codeword are stored on different physical nodes, and an $(n,k)$ MDS code can recover the entire codeword by accessing any $k$ of its coordinates. Coding for distributed storage gives rise to a new set of problems in coding theory related to the need of reducing inter-node communication in the system. A large number of recent papers addressed the problem of optimizing the total amount of information downloaded for repair of a single failed node (the repair bandwidth) by accessing information on $d$ {\em helper nodes}, where $k\le d\le n-1.$

By the so-called cut-set bound (Dimakis et al., 2010), the repair bandwidth of an $(n,k=n-r)$ MDS code using $d$ helper nodes is at least $dl/(d+1-k),$ where $l$ is the size of the node. Also, a number of known constructions of MDS array codes meet this bound with equality. In a related but separate line of work, Guruswami and Wootters (2016) studied repair of Reed-Solomon (RS) codes, showing that these codes can be repaired using a smaller bandwidth than under the trivial approach. At the same time, their work as well as follow-up papers stopped short of constructing RS codes (or any scalar MDS codes) that meet the cut-set bound with equality, which has been an open problem in coding theory.

In this work we present a solution to this problem, constructing RS codes of length $n$ over the field $q^l, l=\exp((1+o(1))n\log n)$ that meet the cut-set bound. We also prove an almost matching lower bound on $l$, showing that the super-exponential scaling is both necessary and sufficient for achieving the cut-set bound \textcolor{black}{using linear repair schemes}. More precisely, we prove that for scalar MDS codes (including the RS codes) to meet this bound, the sub-packetization $l$ must satisfy $l \ge \exp((1+o(1)) k\log k).$ 
\end{abstract}

\section{Introduction}\label{section:repairing RS}

\subsection{Minimum Storage Regenerating codes and optimal repair bandwidth}
The amount of information produced has grown exponentially over the last decade, and large-scale storage systems are widely used to store the data. The problem that we consider is motivated by applications of
codes in distributed storage wherein the data is written on a large number of physical storage nodes. Failure of an individual node
renders a portion of the data inaccessible, and erasure correcting codes are used to increase the reliability of the system. 
The repair task performed by the system relies on communication between individual nodes, and introduces new challenges in the code
design. In particular, a new parameter that has a bearing on the overall efficiency of the system is the amount of data sent between the 
nodes in the process of repair. 

To protect the information, we divide the original file into $k$ information blocks and view each block as a single element of a finite field $F$ or a vector over $F$. We encode the data by adding $r=n-k$ parity blocks (field symbols or vectors) and distributing
the resulting $n$ blocks across $n$ storage nodes. In this paper we deal only with linear codes, so the parity blocks are
formed as linear combinations of the information blocks over $F.$ We use the notation $(n,k)$ to refer to the length and dimension of a
linear code. A well-known class of linear $(n,k)$ Maximum Distance Separable (MDS) codes, studied in this paper, has the favorable property that the original file can be recovered from the content stored on any $k$ nodes, which provides the optimal tradeoff between failure tolerance and storage overhead. 

Before proceeding further, we make a brief remark on the terminology used in the literature devoted to erasure correcting codes for
distributed storage. The coordinates of the codeword are assumed to be stored on different nodes, and by extension are themselves 
referred to as nodes. In practice, single node failure is the most common scenario \cite[Section 6.6]{Rashmi14}, so we will be
interested in the problem of designing codes that efficiently correct (repair) a single erasure (failed node). We assume that the data is encoded with a
code $\cC$ over a finite field $F$ wherein each coordinate of the codeword is either an element of $F$ or an $l$-dimensional vector
over $F$, where $l> 1.$ 
The latter construction, termed {\em array codes} turns out to be better suited to the needs of the repair problem,
as will be apparent in the later part of this section.  To repair a failed node, the system needs to download
the contents from some other nodes ({\em helper nodes}) of the codeword to the processor, and the total amount of the downloaded data is 
called the {\em repair bandwidth}. Coding solutions that support efficient repair are called {\em regenerating codes}, and they
have been a focal point of current research in coding theory following their introduction in \cite{Dimakis10}.

One traditional solution to recover a single node failure in an MDS-coded system is to download the content stored on any $k$ nodes. The 
MDS property guarantees that we can recover the whole file, so we can also recover any single node failure. However, this method is far 
from efficient in the sense that the repair bandwidth that it requires is much larger than {is needed} for the repair of a single 
node. Indeed, by a rather counter-intuitive result of \cite{Dimakis10} it is possible to save on the repair bandwidth by
contacting $d>k$ helper nodes, and the maximum savings are attained when $d$ is the largest possible value, namely $d=n-1$. 

More specifically, suppose that an $(n,k)$ MDS-coded system attempts to repair a failed node by connecting to $d$ helper nodes. In this case, as shown in \cite{Dimakis10}, the total amount of information that needs to be downloaded
to complete the repair task is at least $dl/(d+1-k),$ where $l$ is the size of each node. This lower bound on the repair bandwidth is called the {\em cut-set bound} since it is obtained from the cut-set bound in network information theory \cite{ElGamal81}.
Given $k< d\le n-1,$ an $(n,k)$ MDS code achieving the cut-set bound for the repair of any single failed node from any $d$ helper 
nodes is called an $(n,k)$ {\em minimum storage regenerating} (MSR) code with {\em repair degree} $d$ \cite{Dimakis10}.

The definition of MSR codes, given above in an informal way, will be formalized for a particular subclass of codes known as {\em MDS array codes}. 
An $(n,k)$ MDS array code $\cC$ with {\em sub-packetization} $l$ over a finite field $F$ is formed of
$k$ information nodes and $r=n-k$ parity nodes, where every node is a column vector of length $l$ over $F$ (so $\dim_F (\cC)= kl$).
The MDS property requires that any $k$ nodes of $\cC$ suffice to recover the remaining $r$ nodes of the codeword. Array codes are also called {\em vector codes}, while code families more common to coding theory  
 (such as Reed-Solomon (RS) codes and others) are called {\em scalar codes} in the literature. Clearly, scalar codes correspond to the case $l=1$ of the above definition.

\begin{definition}[Repair bandwidth]
Let $\cC$ be an $(n,k)$ MDS array code with sub-packetization $l$ over a finite field $F$.
 For $i\in\{1,\dots,n\}$ and $\cR\subseteq[n]\setminus\{i\}$ with cardinality $|\cR|\ge k$, define 
$N(\cC,i,\cR)$ as the smallest number of symbols of $F$ one needs to download from the helper nodes $\{c_j:j\in\cR\}$ in order to repair the failed node $c_i$. The \emph{repair bandwidth} of the code $\cC$ with $d$ helper nodes equals 
    $$
    \max_{\substack {i\in [n]\\  \cR\subseteq[n]\setminus\{i\}, |\cR|=d}} N(\cC,i,\cR).
    $$
\end{definition}

We note that the symbols downloaded to repair the node $c_i$ can be some functions of the contents  of the helper
nodes $\{c_j,j\in \cR\}.$

\begin{definition}[Cut-set bound \cite{Dimakis10}] \label{def:csb}
Let $\cC$ be an $(n,k)$ MDS array code with sub-packetization $l$ and let $k\leq d \leq n-1$.
For any $i\in[n]$ and any subset
$\cR\subseteq[n]\setminus\{i\}$ of size $d$ we have the following inequality:
\begin{equation}\label{eq:cutset}
N(\cC,i,\cR)\ge \frac{dl}{d+1-k}.
\end{equation}
 An $(n,k)$ MDS array code with sub-packetization $l$ achieving the cut-set bound \eqref{eq:cutset} for the repair of any single failed node from any $d$ helper nodes is called an $(n,k,d,l)$ {\em MSR array code}. 
 \end{definition}

Several constructions of MSR codes are available in the literature: See 
\cite{Ye16,Ye16a,Goparaju17,Raviv17,Tamo13} for the high-rate regime where $k>n/2$, and \cite{Rashmi11} for the low-rate regime where $k
\le n/2$. Recently the concept of repair bandwidth was extended in \cite{Tamo17} to the problem of correcting errors; this paper also
presented explicit code constructions that support error correction under the minimum possible amount of information downloaded during
the decoding process.

 Due to the limited storage capacity of each node, we would like the sub-packetization $l$ to be as small as possible. However, it is shown in \cite{Goparaju14} that for an $(n,k,d=n-1,l)$ MSR array code,  $l\ge \exp({\sqrt{k/(2r-1)}})$  (i.e., $l$ is exponential in $n$ for fixed $r$ and growing $n$). 
 
 \subsection{Repair schemes for scalar linear MDS codes}
While there has been much research into constructions and properties of MSR codes specifically designed for the repair task, it is
also of interest to study the repair bandwidth of general families of MDS codes, for instance, RS codes.
In \cite{Shanmugam14}, Shanmugam et al. proposed a framework for studying the repair bandwidth of a scalar linear $(n,k)$ MDS code $\cC$ 
over some finite field $E$ (called symbol field below). The idea of \cite{Shanmugam14} is to ``vectorize'' the code construction by considering $\cC$ 
as an array code over some subfield $F$ of $E$. This approach provides a bridge between RS codes and MDS array codes, wherein 
the extension degree $l:=[E:F]$ can be viewed as the value of sub-packetization. The code $\cC$ is viewed as an $(n,k)$ MDS array code with sub-packetization $l$, and the repair bandwidth is
defined exactly in the same way as above. The cut-set bound \eqref{eq:cutset} {and the definition of MSR codes} also apply to this setup.

In this paper we study repair of RS codes, focusing on linear repair schemes, i.e., we assume that the repair operations
are linear over the field  $F.$
Guruswami and Wootters \cite{Guruswami16} gave a characterization for
linear repair schemes of scalar linear MDS codes based on the framework in \cite{Shanmugam14}.
We will use this characterization to prove one of our main results, namely, a lower bound on the sub-packetization, so we recall it below.  Let us start with the definition of the dual code.
\begin{definition}[Dual code]
The dual code of a linear code $\cC \subseteq E^n$ is the linear subspace of $E^n$ defined by
$$
\cC^{\perp}=\big\{x=(x_1,\dots,x_n) \in E^n \big|\sum_{i=1}^n x_i c_i = 0 \quad \forall c=(c_1,\dots c_n)\in\cC \big\}.
$$
\end{definition}
In the next theorem $E$ is the degree-$l$ extension of the field $F$. Viewing $E$ as an $l$-dimensional vector space over $F$, we use the notation
$\dim_F(a_1,a_2,\dots,a_t)$ to refer to the dimension of the subspace spanned by the set $\{a_1,a_2,\dots,a_t\}\subset E$ over $F$.

We will need a result from \cite{Guruswami16} which we state in the form that is suited to our needs.
\begin{theorem}[\cite{Guruswami16}]\label{Thm:Guru}
Let $\cC\subseteq E^n$ be a scalar linear MDS code of length $n$. Let $F$ be a subfield of $E$ such that $[E:F]=l.$ For a given $i\in \{1,\dots,n\}$  the following statements are equivalent.

\begin{enumerate}[(1)]
\item There is a linear repair scheme of the node $c_i$ over $F$ 
such that the repair bandwidth $N(\cC,i,[n]\setminus\{i\}) \le b$.

\item There is a subset of codewords $\cP_i \subseteq \cC^{\perp}$
with size $|\cP_i|=l$ such that
$$
\dim_F ( \{x_i: x\in \cP_i\}) = l,
$$
and 
$$
b \ge \sum_{j\in[n]\setminus\{i\}} \dim_F ( \{x_j: x\in \cP_i\})
$$
\end{enumerate}
\end{theorem}

In addition to this general linear repair scheme for scalar linear MDS codes, the authors of \cite{Guruswami16} also presented a specific repair scheme for a family of RS codes and further proved that (in some cases) the repair bandwidth of 
RS codes using this scheme is the smallest possible among all linear repair schemes and all scalar linear MDS codes with the same 
parameters. 
At the same time, the approach of \cite{Guruswami16} has some limitations. 
Namely, their repair scheme applies only for small sub-packetization $l=\log_{n/r} n$, and the optimality claim only holds for this specific sub-packetization value.  At the same time, in order to achieve the cut-set bound, $l$ needs to be exponentially large in $n$ for a fixed 
value of $r$ \cite{Goparaju14}, so the repair bandwidth of this scheme is rather far from the bound.
Subsequently, Ye and Barg \cite{Ye16b} used the general linear repair scheme in 
\cite{Guruswami16} to construct an explicit family of RS codes with asymptotically optimal repair bandwidth: the ratio between the actual 
repair bandwidth of the codes and the cut-set bound approaches $1$
as the code length $n$ goes to infinity. 

In \cite{Guruswami16}, there is one more restriction on the parameters of the RS codes, namely they achieve the smallest possible
repair bandwidth only if the number of parities is of the form $r=q^s,(l-s)|l.$ In \cite{Dau17}, Dau and Milenkovic generalized the scheme in \cite{Guruswami16} and extended their
results to all values of $s=1,\dots,l-1$. The repair bandwidth attained in \cite{Dau17} is $(n-1)(l-s)$ symbols of $F$
 for $r\ge q^s$, and is the smallest possible whenever $r$ is a power of $q.$ In \cite{Dau16}, Dau et al. extended the results of \cite{Guruswami16} to repair of multiple erasures.

To summarize the earlier work, constructions of RS codes (or any scalar MDS codes) that meet the cut-set bound have as yet been unknown, so the existence question of such codes has been an open problem. In this paper, we resolve this problem in the affirmative, presenting such a construction. We also prove a lower bound on the sub-packetization of scalar linear MDS codes that attain the cut-set bound with a linear repair scheme, showing that there is a penalty for the scalar case compared to MDS array codes.

\subsection{Our Results}\label{sect:results}
\begin{enumerate}[(1)]
\item
{\bf Explicit constructions of RS codes achieving the cut-set bound:} Given any $n,k$ and $d, k\le d\le n-1$, we construct an $(n,k)$ RS code over the field $E=\ff_{q^l}$ that achieves the cut-set bound \eqref{eq:cutset} when repairing {\em any} single failed node from {\em any} $d$ helper nodes. As above, we view RS codes over $E$ as vector codes over the subfield $F=\ff_q$. 
The main novelty in our construction is the choice of the evaluation points for the code in such a way that the degrees of the {evaluation} points over $F$ are distinct primes. As a result, the symbol field is an extension field of $F$ with degree no smaller than the product of these distinct primes. For the actual repair we rely on the linear scheme proposed in \cite{Guruswami16} (this is essentially the only possible linear repair approach).

The value of sub-packetization $l$ of our construction equals $s$ times the product of the first $n$ 
distinct primes in an arithmetic progression, 
  $$
   l= s \biggl(\prod_{\substack{i=1\\[.02in]p_i\equiv 1\text{ mod } s}}^n p_i\biggr),
  $$
  where $s:= d+1-k.$ This product is a well-studied function in number theory, related to a classical arithmetic function $\psi(n,s,a)$  (which is essentially the sum of logarithms of the primes). The prime number theorem in arithmetic progressions (for instance, \cite[p.121]{IK04}) yields asymptotic estimates for $l$. In particular, for fixed $s$ and large $n$, we have $l= e^{(1+o(1)) n\log n}.$ 

In contrast, for the case $d=n-1$ (i.e., $s=r=n-k$), there exist MSR array codes that attain sub-packetization $l=r^{\lceil n/(r+1) \rceil}$ \cite{Wang16},
which is the smallest known value among MSR codes\footnote{ The construction of \cite{Wang16} achieves the cut-set bound only for repair of systematic nodes, and gives $l=r^{\lceil k/(r+1) \rceil}$. Using the approach of \cite{Ye16}, it is possible to modify the construction of \cite{Wang16} and to obtain an MSR code with $l=r^{\lceil n/(r+1) \rceil}$.}.  So although this  distinct prime structure allows us to achieve the cut-set bound, it makes us pay a penalty on the sub-packetization. 

\vspace*{.1in}\item {\bf A lower bound on the sub-packetization of scalar MDS codes achieving the cut-set bound:} 
Surprisingly, we also show that the distinct prime structure discussed above is necessary for any scalar linear MDS code (not just the RS codes) to achieve the cut-set bound under linear repair. Namely, given $d$ such that $k+1\le d\le n-1,$ we prove that for any $(n,k)$ scalar linear MSR code with repair degree
$d,$ 
 the sub-packetization $l$ is bounded below by  $l \geq \prod_{i=1}^{k-1}p_i$, where $p_i$ is the $i$-th smallest prime.  By the Prime Number 
Theorem \cite{IK04}, we obtain the lower asymptotic bound on $l$ of the form
$
l\ge e^{(1+o(1))k \log k}.
$

\vspace*{.1in}\item{\bf Main result:}
In summary, we obtain the following results for the smallest possible sub-packetization of scalar linear MDS codes, including the RS codes, whose repair bandwidth 
achieves the cut-set bound.

\begin{theorem}\label{thm:main} Let $\cC$ be an $(n,k=n-r)$ scalar linear MDS code over the field $E=\ff_{q^l},$ 
and let $d$ be an integer satisfying $k+1\le d\le n-1.$ Suppose that for any single failed node of $\cC$ and any $d$ helper nodes there
is a linear repair scheme over $\ff_q$ that uses the bandwidth $dl/(d+1-k)$ symbols of $\ff_q$, i.e., it achieves the cut-set bound \eqref{eq:cutset}. For a fixed $s=d+1-k$ and $n,k\to\infty$ the following bounds on the smallest possible sub-packetization hold true:
  \begin{equation}\label{eq:main} 
  e^{(1+o(1))k \log k}\le l\le e^{(1+o(1)) n\log n}.
  \end{equation}
For large $s$, we have $l\le 
s \prod\limits_{\substack{i: p_i\equiv 1\,\text{\rm mod}\, s}}^n p_i,$
where the product goes over the first $n$ distinct primes in the arithmetic progression.
\end{theorem}
\begin{remark} The bound on $l$ can be made more explicit 
even for large $s$, and the answer depends on whether we accept the Generalized Riemann Hypothesis (if yes, we can still claim the bound 
$l\le \exp((1+o(1))n\log n)$). \end{remark}

\item{\bf Discussion: Array codes and scalar codes} The lower bound in \eqref{eq:main} is 
much larger than the sub-packetization of many known MSR array code constructions.
To make the comparison between the repair parameters of scalar codes and array codes more clear, we summarize the tradeoff between the repair bandwidth and the sub-packetization of some known MDS code constructions in the following table.
We only list the papers considering the repair of a single node from all the remaining $n-1$ helper nodes. Moreover, in the table we limit ourselves to explicit code constructions, and do not list multiple existence results that appeared in recent years.

\begin{table}
\captionof{table}{Tradeoff between repair bandwidth and sub-packetization}
\begin{center}{\normalsize \begin{tabular}{| c | c | c | c |} 
\hline Code construction
  & Repair bandwidth &  sub-packetization  & achieving cut-set bound \\ \hline\multicolumn{4}{c}{Array codes}\\\hline
\begin{tabular}[c]{@{}c@{}c@{}} $(n,k=n-r,n-1,l)$\\ MSR array codes for \\ $2k\le (n+1)$, \cite{Rashmi11} 
\end{tabular} & $\frac{(n-1)l}{r}$  & $l=r$ & Yes  \\ \hline
\begin{tabular}[c]{@{}c@{}c@{}} $(n,k,n-1,l)$\\ MSR array codes \\ (a modification of \cite{Wang16}) 
\end{tabular} & $\frac{(n-1)l}{r}$  & $l=r^{\lceil n/(r+1) \rceil}$ & Yes  \\ \hline
\begin{tabular}[c]{@{}c@{}c@{}} $(n,k,n-1,l)$  MSR \\ array codes \cite{Ye16a} 

\end{tabular} & $\frac{(n-1)l}{r}$  & $l=r^{\lceil n/r\rceil}$ & Yes  \\ \hline
\begin{tabular}[c]{@{}c@{}c@{}} $(n,k)$ MDS \\ array codes with design \\ parameter $t\ge1$
\cite{Guruswami17}
\end{tabular} & $(1+\frac{1}{t})\frac{(n-1)l}{r}$  & $l=r^t$ & No  \\  \hline \multicolumn{4}{c}{Scalar codes}\\ \hline
\begin{tabular}[c]{@{}c@{}} $(n,k)$ RS code  \cite{Ye16b}
\end{tabular} & $<\frac{(n+1)l}{r}$  & $l=r^n$ & No  \\  \hline
\begin{tabular}[c]{@{}c@{}} $(n,k)$ RS code  \cite{Guruswami16}
\end{tabular} & $n-1$  & $l=\log_{n/r} n$ & No  \\   \hline
\begin{tabular}[c]{@{}c@{}} $(n,k)$ RS code  \cite{Dau17}\\  
\end{tabular} &  $(n-1)l(1-\log_nr)$ & $\log_q n$ & No  \\   \hline
\begin{tabular}[c]{@{}c@{}} $(n,k)$ RS code\\  (this paper)
\end{tabular} & $\frac{(n-1)l}{r}$  & $l\approx n^{ n}$ & Yes  \\ 
\hline
\end{tabular}}
\end{center}
\end{table}

As discussed earlier, the constructions of \cite{Guruswami16,Dau17} have optimal repair bandwidth among all the RS codes with the same sub-packetization value as in these papers\footnote{Expressing the sub-packetization of the construction in \cite{Dau17} via $n$ and $r$ is
difficult. The precise form of the result in \cite{Dau17} is as follows: for every $s<l$ and $r\ge q^s,$ the authors construct repair schemes of RS codes of length $n=q^l$
with repair bandwidth $(n-1)(l-s).$ Moreover, if $r=q^s,$ then the schemes proposed in \cite{Dau17} achieve the smallest possible repair bandwidth for codes with these parameters.}.
At the same time, these values are too small for the constructions of \cite{Guruswami16,Dau17} to achieve the cut-set bound.
From the first three rows of the table one can clearly see that the achievable sub-packetization values for MSR array codes are much smaller than the lower bound for scalar linear MSR codes derived in this paper. This is to be expected since for array codes we only require the code to be linear over the ``repair field,'' i.e., $F$, and not the symbol field $E$ as in the case of scalar codes. 
\end{enumerate}

\subsection{Organization of the paper}
In Sec.~\ref{sect:warmup}, we present a {simple construction} of RS codes  that achieve the cut-set bound for some of the nodes. This
construction is inferior to the more involved construction of Sec.~\ref{Sect:cons}, but simple to follow, and already contains some
of the main ideas of the later part, so we include it as a warm-up for the later results. In Sec.~\ref{Sect:cons}, we
present our main construction of RS codes that achieve the cut-set bound for the repair of any single node, proving the upper estimate in \eqref{eq:main}. In Sec.~\ref{Sect:lb}, we prove the lower bound on the sub-packetization of scalar linear MSR codes, finishing the proof of \eqref{eq:main}.

\section{A simple construction}\label{sect:warmup}
In this section we present a simple construction of RS codes that achieve the cut-set bound for the repair of certain nodes. 
We note that any $(n,k)$ MDS code trivially allows repair that achieves the cut-set bound for $d=k$. We say that a node in an MDS code has a {\em nontrivial optimal repair scheme} if {for a given $d>k$} it is possible to repair this node from any $d$  helper nodes with repair bandwidth achieving the cut-set bound. 
The code family presented in this section is different from standard MSR codes in the sense that although the repair bandwidth of our construction achieves the cut-set bound, the number of helper nodes depends on the node being repaired.

Let us first recall the definition of (generalized) Reed-Solomon codes.
\begin{definition}
A \emph{Generalized Reed-Solomon code} $\text{\rm GRS}_F(n,k,\Omega,v)\subseteq F^n$ of dimension $k$ over $F$ 
with evaluation points $\Omega=\{\omega_1,\omega_2,\dots,\omega_n\}\subseteq F$  is the set of vectors
\begin{align*}
\{(v_1f(\omega_1),\dots,v_nf(\omega_n))\in F^n:f\in F[x], \deg f\le k-1\}
\end{align*}
where $v=(v_1,\dots,v_n)\in (F^\ast)^n$ are some nonzero elements. If $v=(1,\dots,1),$ then the GRS code is called
a Reed-Solomon code and is denoted as $\text{\rm RS}_F(n,k,\Omega)$.

It is well known \cite[p.304]{MacWilliams77} that 
   \begin{equation}\label{eq:grs}
    (\text{\rm RS}_E(n,k,\Omega))^\bot=\text{\rm GRS}_E(n,n-k,\Omega,v)
  \end{equation}
where $v_i=\prod_{j\ne i}(\omega_i-\omega_j)^{-1}, i=1,\dots,n$. (The dual of an RS code is a GRS code.)  
\end{definition}   

Denote by $\pi(t)$ the number of primes less than or equal to $t$. Let $F$ be a finite field and let $E$ be the extension of $F$ of degree
$t.$ The trace function $\trace_{E/F}:E\to F$ is defined by
  $$
   \trace_{E/F}(x):=x+x^{|F|}+x^{|F|^2}+\dots+x^{|F|^{t-1}}.
   $$

In the next theorem we construct a special subfamily of RS codes. Our construction enables nontrivial repair of $\pi(r)$ nodes, which
without loss of generality we take to be nodes $1,2,\dots,\pi(r)$. Let $d_i, i=1,2,\dots,\pi(r)$ be the number of helper nodes used to repair the $i$-th node. We will take $d_i=p_i+k-1$, where $p_i$ is the $i$-th smallest prime number.  The repair scheme presented below
supports repair of node $i$ by connecting to any $d_i$ helper nodes and downloading a $\frac{1}{p_i}$-th
proportion of information stored at each of these nodes. Since $p_i=d_i-k+1,$ this justifies the claim of achieving the cut-set bound for repair of a single node.

\begin{theorem} Let $n\ge k$ be two positive integers, and let $r=n-k.$ 
There exists an $(n,k)$ RS code over a field $E$ such that $\pi(r)$ of its coordinates admit nontrivial optimal repair schemes.
\end{theorem}

\begin{IEEEproof} Let $m:=\pi(r)$ and let $q\ge n-m$ be a prime power. Let $E$ be the $\big(\prod_{i=1}^m p_i\big)$-th degree extension of the finite field $\mathbb{F}_q$. 

Let $\alpha_i, i=1,\dots,m$ be an element of order $p_i$ over $\mathbb{F}_q,$ so that $\mathbb{F}_{q^{p_i}}=\mathbb{F}_q(\alpha_i),$
where $\mathbb{F}_q(\alpha_i)$ denotes the field obtained by adjoining $\alpha_i$ to $\mathbb{F}_q.$ It is clear that $E=
\mathbb{F}_q(\alpha_1,\dots,\alpha_m)$.
Define $m$ subfields $F_i$ of $E$ by setting 
     $$
     F_i=\mathbb{F}_q(\alpha_j:j\neq i),
     $$
 so that $E=F_i(\alpha_i)$ and $[E:F_i]=p_i$, $i=1,\dots,m.$
Let $\alpha_{m+1},\dots,\alpha_{n}\in \mathbb{F}_q$ be arbitrary $n-m$ distinct elements of the field, and let $\Omega=\{\alpha_1,\alpha_2,\dots,\alpha_n\}$.

Let $\cC=\text{\rm RS}_E(n,k,\Omega)$ be the RS code of dimension $k$ with evaluation points $\Omega$ and let $\cC^\bot$ be its dual code.
We claim that for $i=1,2,\dots,m,$ the $i$-th coordinate (node) of $\cC$ can be optimally repaired from any $d_i$ helper nodes, where
$$
d_i= p_i+k-1 .
$$

Let $i\in\{1,2,\dots,m\}$ and let us show how to repair the $i$th node. Choose a subset of helper nodes $\cR_i\subseteq [n]\backslash\{i\},|\cR_i|=d_i,$ and note that since $p_i\leq r,$ we have $d_i\le n-1$.
 Let $h(x)$ be the annihilator polynomial of the set $\{\alpha_j:j\in[n] \setminus(\cR_i \cup \{i\})\}$, i.e., 
\begin{equation}\label{eq:stt}
h(x)=\prod_{j\in[n] \setminus(\cR_i \cup \{i\}) } (x-\alpha_j).
\end{equation}
Since $\deg(h(x))=n-k-p_i,$ we have $\deg(x^s h(x))<r$ for all $s=0,1,\dots,p_i-1.$
As a result, for all $s=0,\dots,p_i-1$, the vector
\begin{equation}\label{eq:ed}
(v_1 \alpha_1^s h(\alpha_1),\dots,v_n \alpha_n^s h(\alpha_n))\in \cC^\bot,
\end{equation}
cf. \eqref{eq:grs}.
Let $c=(c_1,\dots,c_n)\in \cC$ be a codeword. By \eqref{eq:ed} we have
$$
\sum_{j=1}^n v_j h(\alpha_j) \alpha_j^s c_j=0, \quad s=0,\dots,p_i-1.
$$
Let $\trace_i:=\trace_{E/F_i}$ denote the trace from $E$ to $F_i.$ We have
$$
\sum_{j=1}^n\trace_i(v_j h(\alpha_j) \alpha_j^s c_j)=0, \quad s=0,\dots,p_i-1.
$$
Equivalently, we can write
\begin{equation}\label{eq:trg}
\begin{aligned}
\trace_i(v_i h(\alpha_i) \alpha_i^s c_i) & = 
- \sum_{j\neq i}\trace_i(v_j h(\alpha_j) \alpha_j^s c_j) \\
& = - \sum_{j\in \cR_i}\trace_i(v_j h(\alpha_j) \alpha_j^s c_j)  \\
& = - \sum_{j\in \cR_i}  \alpha_j^s \trace_i(v_j h(\alpha_j) c_j),
 \quad s=0,\dots,p_i-1,
\end{aligned}
\end{equation}
where the second equality follows from \eqref{eq:stt} and the third follows because $\alpha_j\in F_i$ for all $j\neq i$ and $\trace_i$ is an $F_i$-linear map.

The information used to recover the value $c_i$ (to repair the $i$th node) is comprised of the following $d_i$ elements of $F_i:$
   $$
   \trace_i(v_j h(\alpha_j) c_j), \;\;j\in\cR_i.
   $$
Let us show that these elements indeed suffice. First, by \eqref{eq:trg}, given these elements, we can calculate
the values of $\trace_{i}(v_i h(\alpha_i) \alpha_i^s c_i)$
for all $s=0,\dots,p_i-1.$ The mapping  
   $$
   \begin{array}{ll}
   E\to F_i^{p_i}\\
\gamma \mapsto \big(\trace_{i} \big(v_ih(\alpha_i)\gamma \big), 
\trace_{i} \big( v_i h(\alpha_i)\alpha_i \gamma \big) , \dots, 
\trace_{i} \big( v_i h(\alpha_i) \alpha_i^{p_i-1} \gamma \big) \big).
  \end{array}
   $$ 
is in fact a bijection, which can be realized as follows. Since the set $\{1,\alpha_i,\dots,\alpha_i^{p_i-1}\}$  forms a basis of $E$ over $F_i$
and $v_i h(\alpha_i) \neq 0,$ the set $\{v_i h(\alpha_i), v_i h(\alpha_i)\alpha_i,\dots, v_i h(\alpha_i)\alpha_i^{p_i-1}\}$
also forms a basis. 
{Let $\{ \theta_0, \theta_1, \dots, \theta_{p_i-1} \}$ be the dual basis of $\{v_i h(\alpha_i), v_i h(\alpha_i)\alpha_i,\dots, v_i h(\alpha_i)\alpha_i^{p_i-1}\},$ i.e., 
$$
\trace_i(v_i h(\alpha_i)\alpha_i^s  \theta_j) = \left\{
\begin{array}{cc}
0, & \text{~if~} s \neq j \\
1, & \text{~if~} s = j 
\end{array} \right. 
\text{for all~} s,j\in\{0,1,\dots,p_i-1\}.
$$
The value $c_i$ can now be found as follows:
$$
c_i= \sum_{s=0}^{p_i-1} \trace_i(v_i h(\alpha_i)\alpha_i^s c_i) \theta_s.
$$}
 (this is the essence of the repair scheme
proposed in \cite{Guruswami16}). 

The presented arguments constitute a linear repair scheme of the node $c_i, i=1,\dots m$ over $F_i.$ 
The information downloaded from each of the helper nodes consists of one element
of $F_i,$ or, in other words, the $(1/p_i)$th proportion of the contents of each node.
This shows that node $i$ admits nontrivial optimal repair. 
The proof is thereby complete.
\end{IEEEproof}

\begin{example} Take $q=5,$ $k=3,r=5$. We have $\pi(r)=3$ and $p_1=2, p_2=3, p_3=5.$ 
Let us construct an $(8,3)$ RS code over the field $E=\mathbb{F}_{5^{30}}$, where the first $3$ nodes admit  nontrivial optimal repair schemes. Let  $\alpha$ be a primitive element of $E$. Choose the set $\Omega=\{\alpha_1,\dots,\alpha_8\}$ as follows:
   $$\alpha_1=\alpha^{\frac{5^{30}-1}{5^2-1}},\alpha_2=\alpha^{\frac{5^{30}-1}{5^3-1}},\alpha_3=\alpha^{\frac{5^{30}-1}{5^5-1}},\alpha_4=0,\alpha_5=1,\alpha_6=2,\alpha_7=3,\alpha_8=4.
   $$
  The number of helper nodes for the first $3$ nodes is $(d_1,d_2,d_3)=(4,5,7)$.
It is easy to verify that for any subset  $A\subseteq \{1,2,3\}$
$$ \mathbb{F}_5(\alpha_i:i\in A)=\mathbb{F}_{m_{\!_A}}, \text{ where } m_{_A}=5^{(\prod_{i\in A}p_i)}.$$

The code $\cC$ constructed in the above proof is given by $\cC=\text{\rm RS}_E(8, 3,\Omega).$ Let us address the task
of repairing $c_3$ from all the remaining $7$ helper nodes with repair bandwidth achieving the cut-set bound.
Let $\cC^\bot=\text{\rm GRS}_E(8, 5,\Omega,v)$, where 
$v=(v_1,\dots,v_8)\in (E^\ast)^8$.
We download the value $\trace_{E/\mathbb{F}_{5^{6}}}(v_jc_j)$ from each helper node $c_j,j\neq 3$.
Since $[E:\mathbb{F}_{5^{6}}]=p_3,$ this amounts to downloading exactly a $1/p_3=(1/5)$-th fraction
of the information stored at each helper node, which achieves the cut-set bound.
The value of $c_3$ can be found from the downloaded information using the following $5$ equations: 
    $$
  \trace_{E/\mathbb{F}_{5^{6}}}(\alpha_3^s v_3c_3) =
 - \sum_{j\neq 3}\trace_{E/\mathbb{F}_{5^{6}}}(\alpha_j^s v_jc_j)=
 - \sum_{j\neq 3}\alpha_j^s\trace_{E/\mathbb{F}_{5^{6}}}(v_jc_j)
, \quad  s=0,\dots,4.
$$ 
Indeed, the downloaded symbols suffice to recover the vector $(\trace_{E/\mathbb{F}_{5^{6}}}(\alpha_3^s v_3c_3),{s=0,\dots,4})$, and
therefore also suffice to repair the symbol $c_3$. 
\end{example}

\section{A family of RS codes achieving the cut-set bound}\label{Sect:cons}

In this section we develop the ideas discussed above and construct  RS codes achieving the cut-set bound with
nontrivial optimal repair of all nodes. More precisely, given any positive integers $k< d \leq n-1$, we explicitly construct an $(n,k)$ RS code $\cC$ achieving the cut-set bound for the repair of {\em any} single node from {\em any} $d$ helper nodes. In other words, $\cC$ is an $(n,k)$ MSR code with repair degree $d$.

Let $\mathbb{F}_p$ be a finite field of prime order (for simplicity we can take $p=2$). Denote  $s:=d-k+1$ and let $p_1,\dots,p_n$ be $n$ distinct primes such that
\begin{equation}\label{eq:pms}
p_i\equiv 1 \;\text{mod}\, s \;\;\text{~for all~} i=1,2,\dots,n.
\end{equation}
According to Dirichlet's theorem, there are infinitely many such primes. For $i=1,\dots,n$, let $\alpha_i$ be an element of 
degree $p_i$ over $\mathbb{F}_p$, i.e., $[\mathbb{F}_p(\alpha_i):\mathbb{F}_p]=p_i$, and define 
\begin{equation}\label{eq:defF}
\mathbb{F}:=\mathbb{F}_p(\alpha_1,\dots,\alpha_n).
\end{equation}
Note that for any subset of indices $A\subseteq [n]$, the field $\mathbb{F}_p(\{\alpha_i:i\in A\})$ is an extension of $\mathbb{F}_p$
of degree $\prod_{i\in A}p_i,$ and in particular, $\mathbb{F}$ has degree $\prod_{i=1}^np_i$ over $\mathbb{F}_p$. 
Next, we define $n$ distinct subfields  $F_i$ of the field $\mathbb{F}$ and one extension field $\mathbb{K}$ of $\mathbb{F}$. 
    \begin{enumerate}
\item
For $i=1,\dots,n$, define $F_i=\mathbb{F}_p(\{\alpha_j:j\neq i\})$. Note that $\mathbb{F}= F_i(\alpha_i)$ and $[\mathbb{F} : F_i]=p_i$.  
\item The field $\mathbb{K}$ is defined to be the degree-$s$ extension of the field $\mathbb{F}$, i.e. there exists an element $\beta\in \mathbb{K}$ of degree $s$ over $\mathbb{F}$ such that $\mathbb{K}=\mathbb{F}(\beta).$
We also have $[\mathbb{K}: F_i]=s p_i$ for all $i$.
    \end{enumerate}

We are ready to construct a family of RS codes that can be optimally repaired for each node. The set $\alpha_1,\dots,\alpha_n$
serves as the set of evaluation points of the code.

The following theorem is the main result of this section.
\begin{theorem} \label{thm1} 
Let $k,n,d$ be any positive integers such that $k< d < n.$ Let $\Omega=\{\alpha_1,\dots,\alpha_n\}$, where $\alpha_i,i=1,\dots,n$ is an element of degree $p_i$
over $\mathbb{F}_p$ and $p_i$ is the $i$th smallest prime that satisfies \eqref{eq:pms}.
 The code $\cC:=\text{\rm RS}_{\mathbb{K}}(n,k,\Omega)$ achieves the cut-set bound for the repair of any single node from any $d$ helper nodes.
In other words, $\cC$ is an $(n,k)$ MSR code with repair degree $d$. 
 \end{theorem}
\textcolor{black}{
\begin{remark} The code constructions in this paper rely on the condition of the form $\alpha_i\not \in \mathbb{F}_q(\alpha_j,j\ne i), i=1,\dots,n$ (in this section we also require that the extension degree $[\mathbb{F}:F_i] \equiv 1 \text{~mod~} s, i=1,\dots,n$).
 The most efficient way to accomplish this in terms of the value of sub-packetization $l$ is to take the extension degrees to be
the smallest (distinct) primes, and this is the underlying idea behind the code constructions presented in this paper.
\end{remark}}

\begin{IEEEproof}
Our repair scheme of the $i$-th node is performed over the field $F_i$. More specifically, for every $i\in[n]$, we explicitly construct a vector space $S_i$ over the field $F_i$ such that
\begin{equation}\label{eq:Si}
\dim_{F_i} S_i = p_i, \quad
  S_i + S_i\alpha_i+\dots + S_i\alpha_i^{s-1}=\mathbb{K},
\end{equation}
where $S_i \alpha := \{\gamma \alpha: \gamma \in S_i\}$,
and the operation $+$ is the Minkowski sum of sets, $T_1 + T_2 := \{\gamma_1+\gamma_2:\gamma_1\in T_1, \gamma_2\in T_2 \}.$ Note that the sum in \eqref{eq:Si} is in fact a direct sum since the dimension of each summand is $p_i$, and $[\mathbb{K}:F_i]=sp_i$.
We will describe a construction of $S_i$ and prove that $S_i$ satisfies \eqref{eq:Si} in Lemma~\ref{lem1} later in this section. For now let us assume that we have such vector spaces $S_i,i=1,2,\dots,n$ and continue the proof of the theorem.

Suppose that we want to repair the $i$-th node from a subset $\cR\subseteq [n]\backslash\{i\}$ of 
$|\cR|=d$ helper nodes. 
Let $h(x)$ be the annihilator polynomial of the set $\{\alpha_j: j\in[n]\setminus (\cR\cup\{i\}) \}$, i.e., 
\begin{equation}\label{eq:defh}
h(x)=\prod_{j\in[n]\setminus (\cR\cup\{i\})}(x-\alpha_j).
\end{equation}
By \eqref{eq:grs} the dual code of $\cC$ is $\cC^\bot=\text{\rm GRS}_{\mathbb{K}}(n,n-k,\Omega,v)$ where the coefficients  
$v=(v_1,\dots,v_n) \in (\mathbb{K}^*)^n$ are nonzero.
Clearly, $\deg(x^th(x))\leq s-1+n-(d+1)<n-k$ for all $t=0,1,\dots,s-1,$ so for any such $t$ we have
\begin{equation}\label{eq:dual}
(v_1\alpha_1^t h(\alpha_1),\dots,v_n \alpha_n^t h(\alpha_n))\in\cC^\bot.
\end{equation}
These $s$ dual codewords will  be used to recover the $i$-th coordinate. 
Let $c=(c_1,\dots,c_n)\in \cC$ be a codeword, and let us construct a repair scheme for the coordinate (node) $c_i$ using
the values $\{c_j:j\in \cR\}$. Rewrite \eqref{eq:dual} as follows:
\begin{equation}\label{eq:inter}
\sum_{j=1}^n  v_j\alpha_j^t h(\alpha_j) c_j =0 \text{~for all~}  t=0,\dots,s-1.
\end{equation}
Let $e_1,\dots,e_{p_i}$ be an arbitrary basis of the subspace $S_i$ over the field $F_i$. From \eqref{eq:inter} we obtain the
following system of $sp_i$ equations:
$$
\sum_{j=1}^n  e_m v_j\alpha_j^t h(\alpha_j) c_j =0,\;\;   t=0,\dots,s-1;
m=1,\dots,p_i.
$$
Let $\trace_i:=\trace_{\mathbb{K}/F_i}$ be the trace map to the subfield $F_i$. From the last set of equations we have
\begin{equation}\label{eq:2}
\sum_{j=1}^n\trace_i(e_mv_j\alpha_j^th(\alpha_j)c_j)=0 \text{~for all~} t=0,\dots,s-1 \text{~and all~} m=1,\dots,p_i,
\end{equation}
Arguing as in \eqref{eq:trg}, let us write \eqref{eq:2} in the following form:
\begin{equation}\label{eq:rcv}
\begin{aligned}
\trace_i(e_m \alpha_i^t v_i h(\alpha_i)c_i) & = - \sum_{j\neq i} \trace_i(e_mv_j\alpha_j^th(\alpha_j)c_j) \\
& = - \sum_{j\in \cR} \trace_i(e_m v_j \alpha_j^t h(\alpha_j)c_j) \\
& = - \sum_{j\in \cR} \alpha_j^t\trace_i(e_m v_j  h(\alpha_j)c_j)
\text{~for all~} t=0,\dots,s-1 \text{~and all~} m=1,\dots,p_i,
\end{aligned}
\end{equation}
where the second equality follows from \eqref{eq:defh} and the third follows from the fact that the trace mapping $\trace_i$ is $F_i$-linear, and that $\alpha_j\in F_i$ for all $j \neq i$.

As before, to recover $c_i$, we download the following $p_i$ symbols of $F_i$ from each helper node  $c_j, j\in \cR$:
\begin{equation}\label{eq:dl}
\trace_i(e_m v_j h(\alpha_j) c_j) \text{~for~} m=1,\dots,p_i.
\end{equation}
These field elements suffice to recover the node $c_i$. Indeed, according to \eqref{eq:rcv}, we can calculate the values of 
$\trace_{i}(e_m \alpha_i^t v_i h(\alpha_i)c_i)$ for all $t=0,\dots,s-1$ and all $m=1,\dots,p_i$ from the set of elements in \eqref{eq:dl}.
By definition, $e_1,\dots,e_{p_i}$ is a basis of the subspace $S_i$ over the field $F_i$.
According to \eqref{eq:Si}, {
 $\mathbb{K} = S_i + S_i\alpha_i+\dots + S_i\alpha_i^{s-1}$.  }
Therefore, the set $\{e_m\alpha_i^t:  t=0,\dots,s-1 ;\, m=1,\dots,p_i\}$ forms a basis of  $\mathbb{K}$ over  $F_i$
and so does the set  $\{e_m\alpha_i^t v_i h(\alpha_i):  t=0,\dots,s-1 ;\, m=1,\dots,p_i\}$ (recall that $v_i\cdot h(\alpha_i)\neq 0$).
Hence the mapping 
 $$
   \begin{array}{ll}
   \mathbb{K}\to F_i^{s p_i}\\
\gamma \mapsto (\trace_{i}(e_m \alpha_i^t v_i h(\alpha_i) \gamma), m=1,\dots,p_i; t=0,\dots,s-1).
  \end{array}
   $$ 
 is a bijection. 
This means that $c_i$ is uniquely determined by the set of values $\{\trace_{i}(e_m \alpha_i^t v_i h(\alpha_i) c_i),m=1,\dots,p_i; t=0,\dots,s-1\}$, validating our repair scheme.

It is also clear that the construction meets the cut-set bound. Indeed, $c_j\in\mathbb{K}$ for all $j$
 and $[\mathbb{K}:F_i]=sp_i,$ so
the amount of information required from each helper node \eqref{eq:dl} is exactly $(1/s)$th fraction of its contents.

This completes the proof of Theorem \ref{thm1}.
\end{IEEEproof}

In the proof above we assumed the existence of the vector space $S_i$ that satisfies \eqref{eq:Si} for all $i\in[n]$.
In the next lemma we construct such a space and establish its properties. 


For a vector space $V$ over a field $F$ and a set of vectors $A=(a_1,\dots,a_l)\subset V$, let $\spun_F(A)=\{\sum_{i=1}^l \gamma_ia_i, \gamma_i\in F\}$ be the span of $A$
 over $F$.

\begin{lemma}\label{lem1} 
Let $\beta$ be a generating element of $\mathbb K$ over $\mathbb{F}=\mathbb{F}_p(\alpha_1,\dots,\alpha_n).$
Given $i\in [n]$, define the following vector spaces over $F_i$:
\begin{gather*}
S_i^{(1)} =\spun_{F_i} \big(\beta^u \alpha_i^{u+qs}, u=0,1,\dots,s-1; q=0,1,\dots,\textstyle{\frac{p_i-1}{s}}-1 \big) \\
S_i^{(2)}  =\spun_{F_i} \Big(\sum_{t=0}^{s - 1}\beta^t \alpha_i^{p_i-1} \Big) \\
S_i   =S_i^{(1)} + S_i^{(2)}.
\end{gather*}
Then 
$$
\dim_{F_i} S_i = p_i, \quad
S_i + S_i\alpha_i+\dots + S_i\alpha_i^{s-1}=\mathbb{K}.
$$
\end{lemma}

\begin{IEEEproof} Let {$K:=S_i + S_i\alpha_i+\dots + S_i\alpha_i^{s-1}$.} If $K=\mathbb{K},$ then $\dim_{F_i} S_i = p_i$
easily follows. Indeed, by definition $\dim_{F_i} S_i \le p_i$. On the other hand, $[\mathbb{K}:F_i]=s p_i$ and
{$K=\mathbb{K}$} together imply that $\dim_{F_i} S_i \ge p_i$. 

Let us prove that $K=\mathbb{K}.$ Clearly $K$ is a vector space over $F_i$, and $K\subseteq \mathbb{K}$. 
Let us show the reverse inclusion, namely that $\mathbb{K}\subseteq K$. To prove this, recall that 
$\mathbb{K}$ is a vector space of dimension $s$ over $\mathbb{F}$ (see \eqref{eq:defF}), and the set $1,\beta,\dots,\beta^{s-1}$
forms a basis, i.e.,  $\mathbb{K}=\oplus_{u=0}^{s-1}\beta^u\mathbb{F}$.
Thus, the lemma will be proved if we show that 
$\beta^u\mathbb{F} \subseteq K$ for all $u=0,1,\dots,s-1.$ 
To prove this inclusion we will use induction on $u$.

For the induction base, let $u=0$. 
In this case, we have $\alpha_i^{qs} \in S_i^{(1)}$ for all $0\leq q<\frac{p_i-1}{s}$.
Therefore $\alpha_i^{qs+j} \in S_i^{(1)} \alpha_i^j$ for all $0\leq q<\frac{p_i-1}{s}$.
As a result, $\alpha_i^{qs+j} \in K$ for all $0\le q<\frac{p_i-1}{s}$ and all $0 \le j \le s-1$.
In other words, 
\begin{equation}\label{eq:u1}
\alpha_i^t \in K ,\; t=0,1,\dots, p_i-2.
\end{equation}

Next we show that also $\alpha_i^{p_i-1} \in K$.
For every $t=1,\dots,s-1$ we have $0\le \lfloor \frac{p_i-1-t}{s} \rfloor <\frac{p_i-1}{s}$.
As a result,
$$
\beta^t \alpha_i^{t+ \lfloor \frac{p_i-1-t}{s} \rfloor s} \in S_i^{(1)}, \;
t=1,\dots,s-1.
$$
We obtain, for each $t=1,\dots,s-1,$
$$
\beta^t \alpha_i^{p_i-1} = 
\beta^t \alpha_i^{t+ \lfloor \frac{p_i-1-t}{s} \rfloor s} 
\alpha_i^{p_i-1-t - \lfloor \frac{p_i-1-t}{s} \rfloor s}
 \in S_i  \alpha_i^{p_i-1-t - \lfloor \frac{p_i-1-t}{s} \rfloor s} \subseteq K.
$$
At the same time,
$$
\sum_{t=0}^{s - 1}\beta^t \alpha_i^{p_i-1} \in S_i^{(2)} \subseteq K.
$$
 The last two statements together imply that
$$
\alpha_i^{p_i-1} = \sum_{t=0}^{s - 1}\beta^t \alpha_i^{p_i-1}
- \sum_{t=1}^{s - 1}\beta^t \alpha_i^{p_i-1}  \in K.
$$
Combining this with \eqref{eq:u1}, we conclude that
$\alpha_i^t \in K$  for all $t=0,1,\dots, p_i-1$.
Recall that $1,\alpha_i,\dots,\alpha_i^{p_i-1}$  is a basis of $\mathbb{F}$ over $F_i$, and that
$K$ is a vector space over $F_i$, so $\mathbb{F} \subseteq K$.
This establishes the induction base.

Now let us fix $u\ge 1$ and let us assume that $\beta^{u'}\mathbb{F} \subseteq K$ for all $u'<u.$ To prove the induction step, we need to show that $\beta^{u}\mathbb{F}\subseteq K$.
Mimicking the argument that led to \eqref{eq:u1}, we can easily show that
\begin{equation}\label{eq:ubeta}
\beta^u\alpha_i^{u+t} \in K,\; t=0,1,\dots, p_i-2.
\end{equation}
Let us show that \eqref{eq:ubeta} is also true for $t=p_i-1,$ i.e., that $\beta^u\alpha_i^{u+p_i-1} \in K$.
For every $1 \le t \le s-1-u$, we have $0\le \lfloor \frac{p_i-1-t}{s} \rfloor <\frac{p_i-1}{s}$.
As a result,
$$
\beta^{u+t} \alpha_i^{u+t+ \lfloor \frac{p_i-1-t}{s} \rfloor s} \in S_i^{(1)},\; t=1,\dots, s-1-u.
$$
Therefore, for all such $t$
\begin{equation}\label{eq:c1}
\beta^{u+t} \alpha_i^{u+p_i-1} = 
\beta^{u+t} \alpha_i^{u+t+ \lfloor \frac{p_i-1-t}{s} \rfloor s} 
\alpha_i^{p_i-1-t - \lfloor \frac{p_i-1-t}{s} \rfloor s} 
  \in S_i  \alpha_i^{p_i-1-t - \lfloor \frac{p_i-1-t}{s} \rfloor s} \subseteq K
\end{equation}
By the induction hypothesis, $\beta^{u'} \mathbb{F} \subseteq K$ for all $u'=0,1,\dots,u-1.$ As a result,
\begin{equation}\label{eq:c2}
\beta^{u'} \alpha_i^{u+p_i-1} \in K,\; u'=0,1,\dots,u-1.
\end{equation}
At the same time,
\begin{equation}\label{eq:c3}
\sum_{t=0}^{s - 1}\beta^t \alpha_i^{u+p_i-1}
= \Big( \sum_{t=0}^{s - 1}\beta^t \alpha_i^{p_i-1} \Big) \alpha_i^u
\in S_i^{(2)} \alpha_i^u \subseteq K.
\end{equation}
Combining \eqref{eq:c1}, \eqref{eq:c2} and \eqref{eq:c3}, we obtain
$$
\beta^u\alpha_i^{u+p_i-1}
= \sum_{t=0}^{s - 1}\beta^t \alpha_i^{u+p_i-1}
- \sum_{u'=0}^{u-1} \beta^{u'} \alpha_i^{u+p_i-1}
- \sum_{t=1}^{s-1-u} \beta^{u+t} \alpha_i^{u+p_i-1}
\in K.
$$
Now on account of \eqref{eq:ubeta} we can conclude that
$\beta^u\alpha_i^{u+t} \in K$  for all $ t =0,1,\dots, p_i-1.$
Therefore, $\beta^u\mathbb{F} \subseteq K$.
This establishes the induction step and completes the proof of the lemma.
\end{IEEEproof}

The value of sub-packetization of the constructed codes is given in the following obvious proposition.

\begin{proposition}
The sub-packetization of our construction is 
$l=[\mathbb{K}:\mathbb{F}_p]=s\prod_{i=1}^n p_i$, where $p_i$'s are the smallest $n$ distinct primes satisfying \eqref{eq:pms}. 
\end{proposition}
The proof follows immediately from the fact that the repair of the $i$-th coordinate is performed over the field $F_i$, so the repair field of our construction is $\cap_{i=1}^n F_i = \mathbb{F}_p$. To estimate the asymptotics of $l$ for $n\to\infty,$ 
recall that our discussion of Dirichlet's prime number theorem in Sec.~\ref{sect:results} above implies that, for fixed $s$, $l= e^{(1+o(1)) n\log n}$. This proves the upper bound in \eqref{eq:main}.

\section{A lower bound on the sub-packetization of scalar linear MSR codes}\label{Sect:lb}
In this section we prove a lower bound on the sub-packetization value $l$ of $(n,k)$ scalar linear MSR codes, which implies that 
$l\ge e^{(1+o(1))k \log k}$. In contrast, for MSR array codes, a much smaller sub-packetization value $l=r^{\lceil n/(r+1) \rceil}$ is achievable \cite{Wang16}. This shows that limiting oneself to scalar linear  codes necessarily leads to a much larger
sub-packetization, and constructing such codes in real storage systems is even less feasible than their array code counterparts. 
The main result of this section is the following theorem:
\begin{theorem}\label{thm2} Let $F=\mathbb{F}_{q}$ and $E=\mathbb{F}_{q^l}$ for a prime power $q$.
{Let $d$ be an integer between $k+1$ and $n-1$.}
Let ${\cC} \subseteq E^n$ be an $(n,k)$ scalar linear MDS code with a linear repair scheme over $F.$ Suppose that the repair bandwidth of the scheme
achieves the cut-set bound with equality for the repair of any single node from any $d$ helper nodes. Then the sub-packetization $l$ is at least  
    $$ l \geq \prod_{i=1}^{k-1}p_i
    $$ 
    where $p_i$ is the $i$-th smallest prime.  
\end{theorem} 

As discussed above in Sec. \ref{sect:results}, this theorem implies the asymptotic lower bound  $l\ge e^{(1+o(1))k \log k}.$

\vspace*{.1in} In the proof of Theorem \ref{thm2}, we will need the following auxiliary lemmas.
\begin{lemma}\label{lem:subfield} {\rm (Subfield criterion \cite[Theorem 2.6]{Lidl94})} Each subfield  of the field $\ff_{p^n}$ is of order $p^m,$ where $m|n.$
For every positive divisor $m$ of $n$ there exists a unique subfield of $\ff_{p^n}$ that contains $p^m$ elements.
\end{lemma}

\begin{lemma}\label{lem2} Let $E$ be an extension field of  $\ff_q$ and let $\alpha_1,\dots,\alpha_n\in E.$ Then
$$[\ff_q(\alpha_1,\dots,\alpha_n):\ff_q]=\lcm(d_1,\dots,d_n),$$
where $d_i=[\ff_q(\alpha_i):\ff_q].$
\end{lemma}
Proof: Obvious.

\begin{lemma}\label{lem:ez} 
Let $a_1,a_2,\dots,a_n\in F^m$ and $b_1,b_2,\dots,b_n\in F^m$
be two sets of vectors over a field $F$, and let $A$ and $B$ denote their spans over $F$. 
Let $c_i=a_i+b_i, i=1,\dots,n$ then
\begin{equation}\label{eq:plus}
\dim_F(c_1,\dots,c_n) \le \dim A + \dim B.
\end{equation}
\end{lemma}
The lemma follows immediately from the fact that, for any two subspaces $A$ and $B$ of a linear space,
\begin{equation}\label{eq:sumcap}
\dim(A+B)+\dim(A\cap B)=\dim A+\dim B.
\end{equation}
%

In the next lemma $\cS_F(\cdot)$ refers to the row space of the matrix argument over the field $F$.

\begin{lemma}\label{lem:use}
Let $E$ be an extension of a finite field $F$ of degree $l$.
Let $A=(a_{i,j})$ be an $m\times n$ matrix over $E$. Then
  \begin{equation}\label{eq:dz}
\dim(\cS_F(A))
\le \sum_{j=1}^n \dim_F(a_{1,j},a_{2,j},\dots,a_{m,j}).
  \end{equation}
Moreover, if \eqref{eq:dz} holds with equality, then for every $\cJ \subseteq [n],$
  \begin{equation}\label{eq:dzcon}
\dim(\cS_F(A_{\cJ}))
= \sum_{j\in\cJ} \dim_F(a_{1,j},a_{2,j},\dots,a_{m,j})
  \end{equation}
where $A_{\cJ}$ is the restriction of $A$ to the columns with indices in $\cJ$.
\end{lemma}
\begin{IEEEproof} Inequality \eqref{eq:dz} is an immediate consequence of Lemma \ref{lem:ez}. Indeed, suppose that $n=2$ and
view the $i$th row of $A$ as the sum of two $2$-dimensional vectors over $E$, namely $(a_{i,1}|0)$ and $(0|a_{i,2}),i=1,\dots,m;$
then \eqref{eq:dz} is the same as \eqref{eq:plus}. The extension to $n>2$ follows by straightforward induction. 

Now let us prove the second part of the claim. Suppose that 
   $$
   \dim(\cS_F(A))= \sum_{j=1}^n \dim_F(a_{1,j},a_{2,j},\dots,a_{m,j}).
   $$
Then for every $\cJ \subseteq [n],$
\begin{align*}
& \sum_{j\in\cJ} \dim_F(a_{1,j},a_{2,j},\dots,a_{m,j})
+ \sum_{j\in\cJ^c} \dim_F(a_{1,j},a_{2,j},\dots,a_{m,j}) \\ 
= & \dim(\cS_F(A))
\le \dim(\cS_F(A_{\cJ})) + \dim(\cS_F(A_{\cJ^c})).
\end{align*}
But according to \eqref{eq:dz},
\begin{align*}
\dim(\cS_F(A_{\cJ})) \le \sum_{j\in\cJ} \dim_F(a_{1,j},a_{2,j},\dots,a_{m,j}), \\
\dim(\cS_F(A_{\cJ^c})) \le \sum_{j\in\cJ^c} \dim_F(a_{1,j},a_{2,j},\dots,a_{m,j}).
\end{align*}
Therefore
$$
\dim(\cS_F(A_{\cJ})) = \sum_{j\in\cJ} \dim_F(a_{1,j},a_{2,j},\dots,a_{m,j}).
$$
This completes the proof of the lemma.
\end{IEEEproof}

Now we are ready to prove Theorem \ref{thm2}.
{}

\vspace*{.1in}{\em Proof of Theorem \ref{thm2}:}
Let $\cC$ be an $(n,k)$ MSR code with repair degree $d.$ By puncturing the code $\cC$ to any $d+1$ coordinates, we obtain a $(d+1,k)$ MSR code with repair degree $d$.
Therefore without loss of generality below we assume that $d=n-1$.

Let $H=[M | I_r]$ be the parity-check matrix of the code ${\cC}$ over $E$, written in systematic form, where $M$ is an $r\times k$ matrix and $I_r$ is the $r\times r$ identity matrix.  Let $h_{ij}$ be the entry of $H$ in position $(i,j).$ Since ${\cC}$ is an MDS code, every square submatrix of $M$ is invertible. In particular, every entry of $M$ is nonzero, so without loss of generality we may assume that $h_{1,j}=1, j=1,2,\dots, k.$
Since $d\ge k+1$, we also have $n\ge k+2$, and therefore $H$ contains at least two rows.

The theorem will follow from the following claim. 

\begin{claim}
 For $j=1,\dots,k-1$ define $\alpha_j := \frac{h_{2,j}}{h_{2,k}}$.  Then for every $j=1,\dots,k-1$, 
  \begin{equation}\label{eq:claim}
\alpha_j\notin \mathbb{F}_q \big( \big\{ \alpha_i:i\in \{1,2,\dots,k-1\}\setminus\{j\} \big\} \big).
  \end{equation}
 In other words,  $\alpha_j$ is not generated by the remaining  $\alpha_i$'s over $\mathbb{F}_q$.
\end{claim}

We first show that this claim indeed implies the theorem.  
Let $d_i=[\mathbb{F}_q(\alpha_i):\mathbb{F}_q]$ be the degree of the field extension generated by 
$\alpha_i$. We prove by contradiction that for all $j=1,2,\dots,k-1$, $d_j$ does not divide 
$\lcm(d_i:i\in \{1,2,\dots,k-1\} \setminus \{j \})$. Suppose the contrary, i.e., that there is a $j$ such that $d_j|\lcm(d_i:i\in \{1,2,\dots,k-1\} \setminus \{j\}).$ According to Lemma~\ref{lem2},
$$
[\mathbb{F}_q \big( \big\{ \alpha_i:i\in \{1,2,\dots,k-1\}\setminus\{j\} \big\} \big) : \mathbb{F}_q]
= \lcm(d_i:i\in \{1,2,\dots,k-1\} \setminus \{j \}).
$$
Then by Lemma \ref{lem:subfield}, there is a subfield 
    \begin{equation}\label{eq:fj}
F_j \subseteq \mathbb{F}_q \big( \big\{ \alpha_i:i\in \{1,2,\dots,k-1\}\setminus\{j\} \big\} \big) 
    \end{equation}
such that $[F_j :\mathbb{F}_q]=d_j$. Notice that $E=\mathbb{F}_{q^l}$ contains all $\alpha_u, u=1,2,\dots,k-1$.
So both $F_j$ and $\mathbb{F}_q(\alpha_j)$ are subfields of $E$, and they have the same order $q^{d_j}$.
Consequently, $\mathbb{F}_q(\alpha_j) = F_j$. Then from \eqref{eq:fj} we conclude that 
$\alpha_j\in\mathbb{F}_q \big( \big\{ \alpha_i:i\in \{1,2,\dots,k-1\}\setminus\{j\} \big\} \big),$ which contradicts 
\eqref{eq:claim}. Thus, our assumption is wrong, and $d_j\not|\,\lcm(d_i:i\in \{1,2,\dots,k-1\} \setminus \{j \})$. 
As an immediate corollary,
$$
l= [E:\mathbb{F}_q] \ge
[\mathbb{F}_q(\{\alpha_i:i=1,\dots,k-1\}):\mathbb{F}_q]=\lcm(d_1,\dots,d_{k-1})\geq \prod_{i=1}^{k-1}p_i.
$$ 
Thus we have shown that this claim indeed implies the theorem.  Now let us prove the claim.

{\bf Proof of the Claim:} Consider the repair of the $j$-th node of the code $\cC$ for some $j\in \{1,2,\dots,k-1\}$.
Since $\cC$ {can be viewed as} an $(n,k,n-1,l)$ MSR code with a linear repair scheme over $\mathbb{F}_q$, node
$c_j$ can be repaired by downloading $(n-1)l/r$ symbols of $\mathbb{F}_q$ from all the remaining nodes $\{c_i:i\in[n]\setminus\{j\}\},$ where $r=n-k.$ Therefore by  Theorem~\ref{Thm:Guru}, there exist $l$ codewords 
$$
(c_{t,1},c_{t,2},\dots,c_{t,n}) \in \cC^\perp, t=1,2,\dots,l
$$ 
such that
\begin{align}
\dim_{\mathbb{F}_q}(c_{1,j},c_{2,j},\dots,c_{l,j})&= l, \text{ and }  \label{eq:21}\\
\sum_{i\neq j} \dim_{\mathbb{F}_q}(c_{1,i},c_{2,i},\dots,c_{l,i})&=\frac{(n-1)l}{r}. \label{eq:212}
\end{align}
Since $H$ is a generator matrix of $\cC^\perp$, for each $t=1,2,\dots,l$ there is a column vector 
$b_t \in E^r$ such that $(c_{t,1},c_{t,2},\dots,c_{t,n})=b_t^T H$.
We define an $l \times r$ matrix $B$ over the field $E$ as $B=[b_1 b_2 \dots b_l]^T$.
We claim that the $\mathbb{F}_q$-rank of the row space of $B$ is $l$. Indeed, assume the contrary, then there  exists a nonzero vector $w\in \mathbb{F}_q^l$ such that $wB=0$. 
Therefore,
$$
wBH=w
\left[ \begin{array}{cccc}
c_{1,1} & c_{1,2} & \dots & c_{1,n} \\
c_{2,1} & c_{2,2} & \dots & c_{2,n} \\
\vdots & \vdots & \vdots & \vdots \\
c_{l,1} & c_{l,2} & \dots & c_{l,n} \\
\end{array} \right]=0.
$$ 
This implies that $w(c_{1,j},c_{2,j},\dots,c_{l,j})^T=0,$
contradicting \eqref{eq:21}. 
Thus we conclude that $B$ has $l$ linearly independent rows over $\mathbb{F}_q$.

Now we want to show that there exists an $l \times l$ invertible matrix $A$ over $\mathbb{F}_q$ such that the matrix $AB$ is an $r\times r$ block-diagonal matrix $\Diag(a_1,\dots,a_r)$, where each block $a_i$ is formed of a column vector of length $\frac{l}{r}$.
In other words, by performing elementary row operations over $\mathbb{F}_q$, $B$ can be transformed into an $r\times r$ block-diagonal matrix $\Diag(a_1,\dots,a_r)$.
Indeed, for $i\in[n]$, let  $h_i$ be the $i$-th column of the matrix $H$, and define 
$$
t_i=\dim_{\mathbb{F}_q}(Bh_i)
= \dim_{\mathbb{F}_q}(c_{1,i},c_{2,i},\dots,c_{l,i}).
$$
By \eqref{eq:212}, we have
\begin{equation} \label{eq:sumt}
\sum_{i\neq j}^n t_i =\frac{(n-1)l}{r}. 
\end{equation}

Since $H$ generates an $(n,r)$ MDS code, for any subset of indices $\cJ \subseteq [n]$ of size $|\cJ|=r$, the matrix $H_{\cJ}$ is of full rank. 
Therefore, the $l\times r$ matrix $BH_{\cJ}$ satisfies the conditions
    \begin{equation}
l=\dim(\cS_{{\mathbb F}_q}(B))=\dim(\cS_{{\mathbb F}_q}(BH_{\cJ}))\leq \sum_{i\in \cJ}\dim_{\mathbb{F}_q}(Bh_i),
\label{eq:22}
\end{equation} 
where the last inequality follows from Lemma~\ref{lem:use}.
Summing both sides of  \eqref{eq:22} over all subsets $\cJ\subseteq [n]\backslash\{j\}$ of size $|\cJ|=r$, we obtain that
\begin{equation}\label{eq:sand}
\begin{aligned}
l \binom{n-1}{r}&\leq \sum_{\substack{\cJ \subseteq [n]\backslash\{j\} \\|\cJ|=r}}\sum_{i\in \cJ}\dim_{\mathbb{F}_q}(Bh_i) \\
&=\binom{n-2}{r-1}\sum_{i\neq j}t_i  \\
&\overset{\eqref{eq:sumt}}{=} \binom{n-2}{r-1}\frac{(n-1) l}{r}  \\
&=l\binom{n-1}{r},
\end{aligned}
\end{equation}
 This implies that the inequality above is in fact an equality, and therefore, on account of \eqref{eq:22} for every subset $\cJ\subseteq [n]\backslash\{j\},|\cJ|=r$ we have
\begin{equation}
l =\sum_{i\in \cJ}\dim_{\mathbb{F}_q}(Bh_i)=\sum_{i\in \cJ}t_i.
\label{eq:222}
\end{equation}
From \eqref{eq:222} we obtain that for all $i\in[n]\setminus\{j\}$ 
\begin{equation}\label{eq:sing}
\dim_{\mathbb{F}_q}(Bh_i)=t_i=l/r.
\end{equation}
Moreover, since \eqref{eq:22} holds with equality, we can use the second part of Lemma~\ref{lem:use} to claim that, for 
$\cJ\subseteq [n]\backslash\{j\}$ of size $|\cJ|\le r,$
    \begin{equation}\label{eq:tem}
\dim(\cS_{\mathbb{F}_q}(BH_{\cJ}))
= \sum_{i\in \cJ}\dim_{\mathbb{F}_q}(Bh_i)
= \frac{|\cJ|l}{r}. 
    \end{equation}
Let us take $\cJ$ to be a subset of $\{k+1,k+2,\dots,n\}.$ 
Since the last $r$ columns of $H$ form an identity matrix, \eqref{eq:tem} becomes
    \begin{equation}\label{eq:subm}
\dim(\cS_{\mathbb{F}_q}(B_{\cJ}))
= \frac{|\cJ|l}{r} 
\text{~for all~} \cJ\subseteq [r] \text{~with size~} |\cJ|\le r.
   \end{equation}

Now we are ready to prove that by performing elementary row operations over $\mathbb{F}_q$, $B$ can be transformed into an $r\times r$ block diagonal matrix $\Diag(a_1,\dots,a_r)$, where each block $a_i$ is a single column vector of length $\frac{l}{r}$. 
We proceed by induction.
More specifically, we prove that for $i=1,2,\dots,r$, we can use elementary row operations over $\mathbb{F}_q$ to transform the first $i$ columns of $B$ into the following form:
$$
\left[
\begin{array}{cccc}
a_1 & 0 & \dots & 0 \\
0 & a_2 & \dots & 0 \\
\vdots & \vdots & \vdots & \vdots \\
0 & 0 & \dots & a_i \\
\mathbf{0} & \mathbf{0} & \dots & \mathbf{0}
\end{array}
\right],
$$
where each $\mathbf{0}$ in the last row of the above matrix is a column vector of length $l(1-\frac{i}{r})$.

Let $i=1.$ According to \eqref{eq:subm}, each column of $B$ has dimension $l/r$ over $\mathbb{F}_q$. Thus the induction base holds trivially.
Now assume that there is an $l \times l$ invertible matrix $A$ over $\mathbb{F}_q$ such that
$$
A B_{[i-1]}=
\left[
\begin{array}{cccc}
a_1 & 0 & \dots & 0 \\
0 & a_2 & \dots & 0 \\
\vdots & \vdots & \vdots & \vdots \\
0 & 0 & \dots & a_{i-1} \\
\mathbf{0} & \mathbf{0} & \dots & \mathbf{0}
\end{array}
\right],
$$
where each $\mathbf{0}$ in the last row of this matrix is a column vector of length $l(1-\frac{i-1}{r})$.
Let us write the $i$-th column of $AB$ as $(v_1,v_2,\dots,v_l)^T$. 
Since each column of $B$ has dimension $l/r$ over $\mathbb{F}_q$,
$(v_1,v_2,\dots,v_l)^T$ also has dimension $l/r$ over $\mathbb{F}_q$.
Since the last $l(1-\frac{i-1}{r})$ rows of the matrix $A B_{[i-1]}$ are all zero, we can easily deduce that 
   $$
\dim(\cS_{\mathbb{F}_q}(AB_{[i]})) \le \frac{i-1}{r}l + 
\dim_{\mathbb{F}_q}(v_{(i-1)l/r+1}, v_{(i-1)l/r+2}, \dots, v_l).
   $$
By \eqref{eq:subm}, $\dim(\cS_{\mathbb{F}_q}(AB_{[i]}))=\dim(\cS_{\mathbb{F}_q}(B_{[i]}))=\frac{il}{r}$. 
As a result, 
$$
\dim_{\mathbb{F}_q}(v_{(i-1)l/r+1}, v_{(i-1)l/r+2}, \dots, v_l) \ge l/r = 
\dim_{\mathbb{F}_q}(v_1,v_2, \dots, v_l).
$$
In other words, $(v_{(i-1)l/r+1}, v_{(i-1)l/r+2}, \dots, v_l)$ contains a basis of the set $(v_1,v_2, \dots, v_l)$ over $\mathbb{F}_q$.
This implies that we can use elementary row operations on the matrix $AB$ to eliminate all the nonzero entries
$v_m$ for $m\le (i-1)l/r$, and thus obtain the desired block-diagonal structure for the first $i$ columns.
This establishes the induction step.

We conclude that there exists an $l \times l$ invertible matrix $A$ over $\mathbb{F}_q$ such that  $AB=\Diag(a_1,\dots,a_r)$, where each block $a_i$ is a single column vector of length $\frac{l}{r}$.
For $u\in[r]$, let $A_u$ be the vector space spanned by the entries of $a_u$ over $\mathbb{F}_q$.
According to \eqref{eq:sing}, for all $i\in[n]\setminus\{j\}$
$$
\dim_{\mathbb{F}_q}(ABh_i)=\dim_{\mathbb{F}_q}(Bh_i)=l/r.
$$
Since 
\begin{align*}
\dim_{\mathbb{F}_q}(ABh_i)&=\dim_{\mathbb{F}_q}(\Diag(a_1,\dots,a_r) h_i)\\
&=
\dim_{\mathbb{F}_q}(A_1h_{1,i}+\dots+ A_r h_{r,i}), \quad i=1,2,\dots,n,
\end{align*}
for all $i\in[n]\setminus\{j\}$ we have 
$$
\dim_{\mathbb{F}_q}(A_1h_{1,i}+\dots+ A_r h_{r,i}) =l/r.
$$
Since each column of $B$ has dimension $l/r$ over $\mathbb{F}_q$,
$A_u$ also has dimension $l/r$ over $\mathbb{F}_q$ for every $u\in[r]$.
Recall that $h_{u,i}\neq 0$ for all $u\in[r]$ and all $i\in[k]$.
Thus
$$
\dim_{\mathbb{F}_q}(A_u h_{u,i})=l/r
= \dim_{\mathbb{F}_q}(A_1h_{1,i}+\dots+ A_r h_{r,i})
$$
for all  $u=1,\dots,r$ and $i\in[k]\setminus\{j\}.$
Therefore,
$$
A_1 h_{1,i}=A_2 h_{2,i} = \dots = A_r h_{r,i} \text{~and all~} i\in[k]\setminus\{j\}.
$$
Since $h_{1,i}=1$ for all $i=1,2,\dots,k$, we have
\begin{equation}\label{eq:eqspa}
A_2 h_{2,i} = A_1 \text{~for all~} i\in[k]\setminus\{j\}.
\end{equation}
Equivalently,
$$
A_2 \alpha_i = A_2 \text{~for all~} i\in\{1,2,\dots,k-1\} \setminus\{j\}.
$$
By definition $A_2$ is a vector space over $\mathbb{F}_q$, so 
\begin{equation}\label{eq:stab}
A_2 \gamma = A_2 \text{~for all~} 
\gamma \in \mathbb{F}_q(\{\alpha_i: i\in\{1,2,\dots,k-1\} \setminus\{j\}\}).
\end{equation}
On the other hand,
\begin{equation}\label{eq:cond1}
\begin{aligned}
\dim_{\mathbb{F}_q}(A_1h_{1,j}+\dots+ A_r h_{r,j}) & =\dim_{\mathbb{F}_q}(\Diag(a_1,\dots,a_r) h_j)=
\dim_{\mathbb{F}_q}(ABh_j)\\
& = \dim_{\mathbb{F}_q}(Bh_j) = 
\dim_{\mathbb{F}_q}\{c_{1,j},c_{2,j},\dots,c_{l,j}\}= l,
\end{aligned}
\end{equation}
while
\begin{equation}\label{eq:cond2}
\dim_{\mathbb{F}_q}(A_u h_{u,j})=l/r, \quad  u=1,2,\dots,r.
\end{equation}
Equations \eqref{eq:cond1} and \eqref{eq:cond2} together imply that the vector spaces $A_1 h_{1,j}, A_2 h_{2,j}, \dots, A_r h_{r,j}$ are pairwise disjoint. In particular, $A_1 \cap A_2 h_{2,j} = {0}.$ On account of
\eqref{eq:eqspa}, we therefore have $A_2 h_{2,k} \cap A_2 h_{2,j} = {0}.$
This implies that $A_2 \alpha_j \neq A_2$. By \eqref{eq:stab}, we conclude that
$\alpha_j \notin \mathbb{F}_q(\{\alpha_i: i\in\{1,2,\dots,k-1\} \setminus\{j\}\})$.
This completes the proof of the claim.

\bibliographystyle{IEEEtran}
\bibliography{repair}

\begin{thebibliography}{10}
\providecommand{\url}[1]{#1}
\csname url@samestyle\endcsname
\providecommand{\newblock}{\relax}
\providecommand{\bibinfo}[2]{#2}
\providecommand{\BIBentrySTDinterwordspacing}{\spaceskip=0pt\relax}
\providecommand{\BIBentryALTinterwordstretchfactor}{4}
\providecommand{\BIBentryALTinterwordspacing}{\spaceskip=\fontdimen2\font plus
\BIBentryALTinterwordstretchfactor\fontdimen3\font minus
  \fontdimen4\font\relax}
\providecommand{\BIBforeignlanguage}[2]{{%
\expandafter\ifx\csname l@#1\endcsname\relax
\typeout{** WARNING: IEEEtran.bst: No hyphenation pattern has been}%
\typeout{** loaded for the language `#1'. Using the pattern for}%
\typeout{** the default language instead.}%
\else
\language=\csname l@#1\endcsname
\fi
#2}}
\providecommand{\BIBdecl}{\relax}
\BIBdecl

\bibitem{Rashmi14}
K.~V. Rashmi, N.~B. Shah, D.~Gu, H.~Kuang, D.~Borthakur, and K.~Ramchandran,
  ``A {H}itchhiker's guide to fast and efficient data reconstruction in
  erasure-coded data centers,'' in \emph{ACM SIGCOMM Computer Communication
  Review}, vol.~44, no.~4.\hskip 1em plus 0.5em minus 0.4em\relax ACM, 2014,
  pp. 331--342.

\bibitem{Dimakis10}
A.~G. Dimakis, P.~B. Godfrey, Y.~Wu, M.~J. Wainwright, and K.~Ramchandran,
  ``Network coding for distributed storage systems,'' \emph{IEEE Trans. Inform.
  Theory}, vol.~56, no.~9, pp. 4539--4551, 2010.

\bibitem{ElGamal81}
A.~El~Gamal, ``On information flow in relay networks,'' in \emph{Proc. IEEE
  National Telecom Conf.}, vol.~2, New Orleans, LA, 1981, pp. D4.1.1--D4.1.4.

\bibitem{Ye16}
M.~Ye and A.~Barg, ``Explicit constructions of high-rate {MDS} array codes with
  optimal repair bandwidth,'' \emph{IEEE Transactions on Information Theory},
  vol.~63, no.~4, pp. 2001--2014, 2017.

\bibitem{Ye16a}
------, ``Explicit constructions of optimal-access {MDS} codes with nearly
  optimal sub-packetization,'' 2016, arXiv:1605.08630.

\bibitem{Goparaju17}
S.~Goparaju, A.~Fazeli, and A.~Vardy, ``Minimum storage regenerating codes for
  all parameters,'' \emph{IEEE Transactions on Information Theory}, vol.~63,
  2017.

\bibitem{Raviv17}
N.~Raviv, N.~Silberstein, and T.~Etzion, ``Constructions of high-rate minimum
  storage regenerating codes over small fields,'' \emph{IEEE Transactions on
  Information Theory}, vol.~63, no.~4, pp. 2015--2038, 2017.

\bibitem{Tamo13}
I.~Tamo, Z.~Wang, and J.~Bruck, ``Zigzag codes: {MDS} array codes with optimal
  rebuilding,'' \emph{IEEE Transactions on Information Theory}, vol.~59, no.~3,
  pp. 1597--1616, 2013.

\bibitem{Rashmi11}
K.~V. Rashmi, N.~B. Shah, and P.~V. Kumar, ``Optimal exact-regenerating codes
  for distributed storage at the {MSR} and {MBR} points via a product-matrix
  construction,'' \emph{IEEE Trans. Inform. Theory}, vol.~57, no.~8, pp.
  5227--5239, 2011.

\bibitem{Tamo17}
I.~Tamo, M.~Ye, and A.~Barg, ``Fractional decoding: {E}rror correction from
  partial information,'' 2017, arXiv:1701.06969.

\bibitem{Goparaju14}
S.~Goparaju, I.~Tamo, and R.~Calderbank, ``An improved sub-packetization bound
  for minimum storage regenerating codes,'' \emph{IEEE Transactions on
  Information Theory}, vol.~60, no.~5, pp. 2770--2779, 2014.

\bibitem{Shanmugam14}
K.~Shanmugam, D.~S. Papailiopoulos, A.~G. Dimakis, and G.~Caire, ``A repair
  framework for scalar {MDS} codes,'' \emph{IEEE Journal on Selected Areas in
  Communications}, vol.~32, no.~5, pp. 998--1007, 2014.

\bibitem{Guruswami16}
V.~Guruswami and M.~Wootters, ``Repairing {R}eed-{S}olomon codes,'' \emph{IEEE
  Transactions on Information Theory}, vol.~63, 2017.

\bibitem{Ye16b}
M.~Ye and A.~Barg, ``Explicit constructions of {MDS} array codes and {RS} codes
  with optimal repair bandwidth,'' in \emph{2016 IEEE International Symposium
  on Information Theory (ISIT)}.\hskip 1em plus 0.5em minus 0.4em\relax IEEE,
  2016, pp. 1202--1206.

\bibitem{Dau17}
H.~Dau and O.~Milenkovic, ``Optimal repair schemes for some families of
  full-length {R}eed-{S}olomon codes,'' 2017, arXiv:1701.04120.

\bibitem{Dau16}
H.~Dau, I.~Duursma, H.~M. Kiah, and O.~Milenkovic, ``Repairing {R}eed-{S}olomon
  codes with multiple erasures,'' 2016, arXiv:1612.01361.

\bibitem{IK04}
H.~Iwaniec and E.~Kowalski, \emph{Analytic number theory}.\hskip 1em plus 0.5em
  minus 0.4em\relax American Mathematical Society Providence, RI, 2004,
  vol.~53.

\bibitem{Wang16}
Z.~Wang, I.~Tamo, and J.~Bruck, ``Explicit minimum storage regenerating
  codes,'' \emph{IEEE Transactions on Information Theory}, vol.~62, no.~8, pp.
  4466--4480, 2016.

\bibitem{Guruswami17}
V.~Guruswami and A.~S. Rawat, ``{MDS} code constructions with small
  sub-packetization and near-optimal repair bandwidth,'' in \emph{Proceedings
  of the Twenty-Eighth Annual ACM-SIAM Symposium on Discrete Algorithms}.\hskip
  1em plus 0.5em minus 0.4em\relax SIAM, 2017, pp. 2109--2122.

\bibitem{MacWilliams77}
F.~J. MacWilliams and N.~J.~A. Sloane, \emph{The Theory of Error-Correcting
  Codes}.\hskip 1em plus 0.5em minus 0.4em\relax Elsevier, 1977.

\bibitem{Lidl94}
R.~Lidl and H.~Niederreiter, \emph{Introduction to Finite Fields and Their
  Applications}.\hskip 1em plus 0.5em minus 0.4em\relax Cambridge University
  Press, 1994.

\end{thebibliography}

\end{document}